\documentclass[onecolumn]{IEEEtran}
\usepackage{amsmath}
\usepackage{subcaption}
\usepackage{graphicx}
\usepackage{txfonts}
\usepackage{hyperref}

\usepackage[T1]{fontenc}
\usepackage[utf8]{inputenc}
\usepackage{authblk}

\title{Deep learning forecasts of cosmic acceleration parameters from DECi-hertz Interferometer Gravitational-wave Observatory}

\author[1]{Meng-Fei Sun}
\author[1]{Jin Li \thanks{Corresponding author: cqujinli1983@cqu.edu.cn}}
\author[2,3]{Shuo Cao}
\author[2,3]{Xiaolin Liu}
\affil[1]{College of Physics, Chongqing University, Chongqing 401331, China}
\affil[2]{Institute for Frontiers in Astronomy and Astrophysics, Beijing Normal University, Beijing 102206, China}
\affil[3]{Department of Astronomy, Beijing Normal University, Beijing 100875, China}

\begin{document}

\maketitle % 这里放置标题页

\begin{abstract}
Validating the accelerating expansion of the universe is an important aspect in improving our understanding of the evolution of the universe. By constraining the cosmic acceleration parameter $X_H$, we can discriminate between the cosmological constant plus cold dark matter ($\Lambda \mathrm{CDM}$) model and the Lemaître-Tolman-Bondi (LTB) model.In this paper, we explore the possibility of constraining the cosmic acceleration parameter with the inspiral gravitational waveform of neutron star binaries (NSBs) in the frequency range of 0.1Hz-10Hz, which can be detected by the second-generation space-based gravitational wave detector DECIGO.We used a convolutional neural network (CNN) and a long short-term memory (LSTM) network combined with a gated recurrent unit (GRU), along with a Fisher information matrix to derive constraints on the cosmic acceleration parameter, $X_H$.We assumed that our networks estimate the cosmic acceleration parameter without biases (the expected value of the estimation is equal to the true value). Under this assumption, based on the simulated gravitational wave data with a time duration of one month, we conclude that CNN can limit the relative error to 15.71\%, while LSTM network combined with GRU can limit the relative error to 14.14\%. Additionally, using a Fisher information matrix for gravitational wave data with a five-year observation can limit the relative error to 32.94\%.Under the assumption of an unbiased estimation, the neural networks can offer a high-precision estimation of the cosmic acceleration parameter at different redshifts. Therefore, DECIGO is expected to provide direct measurements of the acceleration of the universe by observing the chirp signals of coalescing binary neutron stars.
\end{abstract}

\section{Introduction}
The verification  of the acceleration of the cosmic expansion is a crucial subject in current astrophysical research. Cosmic acceleration refers to the phenomenon of the universe expanding at an increasingly fast rate and measuring its acceleration is essential for determining the ultimate fate of the universe. Observing the type-Ia Supernovae has provided evidence to support the accelerating expansion of the universe, focusing on a distinctive correlation between the luminosity and distance of supernovae within a specific distance range \cite{1,29,2}. Such a conclusion was further verified by the observations of cosmic microwave background radiation \cite{3} based on the results from Wilkinson Microwave Anisotropy Probe (WMAP) program \cite{4} and Planck collaboration \cite{75}. The analysis of various observational data, including baryon acoustic oscillation, Hubble parameters derived from passively evolving galaxies \cite{77,78,79}, strong gravitational lensing systems \cite{80,81,82,84}, and quasars calibrated as standard rulers and standard candles \cite{85,87} has also suggested that the present universe is undergoing an accelerated phase of expansion. We refer to \cite{5,40} for the summary of recent observational progress made on such issue. Currently, several theoretical models have been proposed to explain the expansion of the universe. Dark energy and modified gravity theories are widely accepted as explanations for these observational facts. Dark energy's significant influence on the evolution of the universe has played a pivotal role in its widespread acceptance within the scientific community. However, assuming that we are at the center of the universe and that the universe is no longer isotropic on large scales and becomes non-uniform can explain the current observational results without using a dark energy model or modifying gravity theory \cite{30,31,32,33,34,35,36,37,38,39,76}. Nevertheless, such an assumption clearly violates the Copernican principle. Therefore, directly detecting the acceleration of the universe's expansion is an important means of verifying the current mainstream theory.

According to the standard cosmological model, the acceleration of the universe's expansion leads to redshift drift, a crucial phenomenon that provides valuable information for understanding the evolution of the universe. Several methods can be used to observe redshift drift. One method involves using Type Ia supernovae as standard candles \cite{6,7}. Another method involves using observations and calculations of cosmic microwave background radiation \cite{8,9}. In this study, we analyze the cosmic acceleration parameter by examining the redshift evolution resulting from cosmic accelerating expansion and the corresponding phase shift of gravitational waves. Such a phase shift caused by the cosmic acceleration was demonstrated by \cite{26}, through a decade of observing gravitational wave signals generated by binary neutron stars. In this analysis, we chose the sensitive frequency range of 0.1Hz-10Hz for the DECIGO (Deci-hertz Interferometer Gravitational Wave Observatory) detector to capture gravitational waveforms generated during the inspiral phase of binary neutron stars. \cite{10} used the covariance matrix derived from the Fisher information matrix to estimate the uncertainty of cosmic acceleration parameter. However, we should note that the phase shift in the low-frequency part of the waveform can be easily overwhelmed by complex noise backgrounds. Therefore, a high-precision parameter estimation method is still required to obtain accurate results. Luckily, the deployment of machine learning algorithms in astronomy has demonstrated its efficacy in accelerating data processing and improving statistical inference. Specially, deep learning in gravitational wave data analysis is becoming crucial to quick and accurate estimation of parameters of interest \cite{11}. In this paper we discuss the ability of different deep learning algorithms,
a convolutional neural network (CNN), and a long short-term memory (LSTM) network combined with a gated recurrent unit (GRU) to measure the cosmic acceleration parameters, based on the time-series waveform of gravitational wave during the binary neutron star inspiral phase in DECIGO. Our results reveal that deep learning is able to provide measurements of the cosmic acceleration parameter with high precision.

The paper is structured as follows. Section 2 is dedicated to the framework of the GW simulations produced for our analysis. Section 3 explains how the estimation of the cosmic acceleration parameter is performed, along with the deep learning results with different deep learning methods. Section 4 presents the results with Fisher information matrix for comparison. Finally, our main conclusions and final remarks are presented in Section 5.

%__________________________________________________________________

\section{Data simulation}

\subsection{Simulation of a gravitational wave signal}

In this paper, we use natural units with $c=G=1$ and select binary neutron stars as our sources of gravitational waves. Our simulation is based on the flat $\Lambda \mathrm{CDM}$ model, with the matter density parameter, $\Omega_{M}=0.3$, and the Hubble constant, $H_0=70$ km/s/Mpc. The relation between redshift drift and the cosmic acceleration parameter is parameterized as \cite{26}: $X_{H}=X(z)/H_{0}$, and $X(z)= \frac{H_{0}}{2}\left(1-\frac{H(z)}{(1+z) H_{0}}\right)$, with the Hubble parameter being $H(z) = H_{0} \sqrt{\Omega_{M}(1+z)^{3}+1-\Omega_{M}}$.

The gravitational wave waveform in the presence of cosmic acceleration expansion is 
\cite{10,26,42}:
\begin{equation}
	\tilde{h}(f)=\left.e^{\Psi_{\text {acc }}(f)} \tilde{h}(f)\right|_{\text {no accel }},
\end{equation}
where the acceleration phase $\Psi_\text {acc }(f)$ is derived from: 
\begin{equation}
	\Psi_{\text {acc }}(f)=-\Psi_{N}(f) \frac{25}{768} X\left(z_{c}\right) \mathcal{M}_{z} x^{4},
\end{equation}
the waveform of stationary phase approximation without acceleration is given by \cite{27,41}: 
\begin{equation}
	\left.\tilde{h}(f)\right|_{\text {no accel }}=\frac{\sqrt{3}}{2} \mathcal{A} f^{-7 / 6} e^{i \psi(f)},
\end{equation}
\begin{equation}
	\mathcal{A}=\frac{1}{\sqrt{30} \pi^{2 / 3}} \frac{\mathcal{M}^{5 / 6}}{D_{L}},
\end{equation}
where $x =\left(\pi \mathcal{M}_{z} \mathit{f}\right)^{2/3}$, $\Psi_{N}(f) = \frac{3}{128}\left(\pi \mathcal{M}_{z} \mathit{f}\right)^{-5/3}$, $\mathcal{M}_{z} = \mathit{M}\left(1+z_{c}\right) \eta^{3/5}$ represents the chirp mass with redshift drift, $z_{c}$ represents the redshift taking into account the cosmic acceleration expansion. 
The chirp mass is defined as $\mathcal{M} = M \eta^{3 / 5}$, with the symmetric mass ratio $\eta=m_{1} m_{2} / M^{2}$ and the total mass $\mathit{M}=m_{1}+m_{2}$. In the case of a flat universe, the luminosity distance to the source is: 
\begin{equation}
	\begin{aligned}
		D_{L}=\frac{1+z}{H_{0}} \int_{0}^{z} \frac{d 	z^{\prime}}{\left[\Omega_{M}\left(1+z^{\prime}\right)^{3}+\Omega_{\Lambda}\right]^{1 / 2}}.
	\end{aligned}
\end{equation}

We used the second-order standard post-Newtonian approximation method to construct the gravitational waveforms \cite{27}:
\begin{equation}
	\begin{aligned}
		\psi(f)= & 2 \pi f t_{c}-\phi_{c}+\frac{3}{128}(\pi \mathcal{M} f)^{-5 / 3}\left\{1+\left(\frac{3715}{756}+\frac{55}{9} \eta\right) \eta^{-2 / 5}\right.\\
		& \times(\pi \mathcal{M} f)^{2 / 3}-16 \pi \eta^{-3 / 5}(\pi \mathcal{M} f)+4 \beta \eta^{-3 / 5}(\pi \mathcal{M} f) \\
		&+\left( \frac{15293365}{508032}
		+ \frac{27145}{504} \eta + \frac{3085}{72} \eta^{2} \right) \eta^{-4/5} (\pi \mathcal{M} f)^{4/3} \\
		&- 10 \sigma \eta^{-4/5} (\pi \mathcal{M} f)^{4/3}\left.\vphantom{\frac{3715}{756}}\right\},
	\end{aligned}
\end{equation}
in the above phase expression, the first term includes the merger time, $t_{c}$, the second term includes the phase at the merger, $\phi_{c}$, and the factor in the parentheses is the standard phase for quadrupole radiation in general relativity. The terms in the parentheses are the expansion terms in the post-Newtonian approximation; then, $|\beta| \lesssim 9.4$ and $|\sigma|\lesssim 2.5$ represent the contributions from spin-orbit coupling and spin-spin coupling to the phase, respectively \cite{43}.

Due to the time-domain signals detected by DECIGO, time-domain signals are more direct and convenient in terms of representation and processing, without the need for additional transformations or processing steps. Furthermore, time-domain signal processing is typically faster than frequency-domain signal processing, which is crucial for handling a large volume of gravitational wave data. In addition, the direct information contained in time-domain signals includes important features and information of gravitational wave signals, such as duration, amplitude, phase, and shape, which are vital for the identification and classification of gravitational wave signals. Moreover, preprocessing and data processing of time-domain signals are relatively simple, allowing the utilization of various filtering techniques, noise reduction methods, and data cleaning techniques to improve the signal-to-noise ratio (S/N) and extract useful features. It is necessary to impose an inverse Fourier transform on the frequency-domain gravitational wave to the time-domain. To simplify the  calculations, we employed the average response function $R_{\text {DECIGO}}$ of the DECIGO detector \cite{44} as:
\begin{equation}
	\begin{aligned}
		R_{\text {DECIGO }} \approx \frac{3}{10} \frac{1}{1+\left(f / f_{*}\right)^{2} /\left[1.85-0.58 \cos \left(2 f / f_{*}\right)\right]},
	\end{aligned}
\end{equation}
where $f /f_{*}=\frac{2 \pi f L}{c}$, $L=1000$ km is the arm length of the DECIGO detector
\cite{45} and $c$ is the speed of light. Considering that our frequency-domain data is obtained from numerical simulations, we use the 1D discrete inverse Fourier transform,
\begin{equation}
	\begin{aligned}
		h(t) &= \frac{1}{2N} \sum_{f=0}^{N-1} e^{i 2\pi f \frac{t}{N}} \tilde{h}(f)R_{\text {DECIGO }}(f),
	\end{aligned}
\end{equation}
where $h(t)$ is the GW strain in the time-domain signal, $N$ is the sample number, and  $\tilde{h}(f)$ is the GW strain in the frequency-domain, $R_{\text {DECIGO}}$(f) represents the average response function, reflecting the detector's sensitivity to signals in different frequencies.

\subsection{Simulation of noise data}

The one-sided noise power spectral density for DECIGO is given by \cite{15,16}: 
\begin{equation}
	\begin{aligned}
	S_{h}^{\text {DECIGO }}(f) & =7.05 \times 10^{-48}\left[1+\left(\frac{f}{f_{p}}\right)^{2}\right]+4.8 \times 10^{-51}\left(\frac{f}{1 H z}\right)^{-4} \\
	& \times\frac{1}{1+\left(\frac{f}{f_{p}}\right)^{2}}+5.53 \times 10^{-52}\left(\frac{f}{1 H z}\right)^{-4} Hz^{-1},
	\end{aligned}
\end{equation}
where $f_{p}=7.36 Hz$. The three terms correspond to short noise, radiation pressure noise, and acceleration noise, respectively. Additionally, we need to consider confusion noise from the galactic and extragalactic white dwarfs \cite{47,48}:
\begin{equation}
	\begin{aligned}
		S_{h}^{\text {gal }}(f) & =2.1 \times 10^{-45}\left(\frac{f}{1 Hz}\right)^{-7 / 3} H z^{-1}, 
	\end{aligned}
\end{equation}
\begin{equation}
	\begin{aligned}
		S_{h}^{\text {exgal }}(f) & =4.2 \times 10^{-47}\left(\frac{f}{1 Hz}\right)^{-7 / 3} Hz^{-1},
	\end{aligned}
\end{equation}
the noise from the galactic and extragalactic white dwarfs is multiplied by a factor of $\mathcal{F} = \exp \left\{-2(f / 0.05 Hz)^{2}\right\}$, which corresponds to the high-frequency cutoff. We also considered the superposition of gravitational waves from many binary neutron star systems, which contribute to the background noise as \cite{16,17}:
\begin{equation}
	\begin{aligned}
		S_{h}^{N S}(f)=1.3 \times 10^{-48}\left(\frac{f}{1 Hz}\right)^{-7 / 3} Hz^{-1}.
	\end{aligned}
\end{equation}

Given the overall noise sources, we could obtain the unilateral noise power spectral density (PSD) of DECIGO as \cite{43}:
\begin{equation}
	\begin{aligned}
		S_{h, \text { DECIGO }}(f)= & \min \left[\frac{S_{h, \text { DECIGO }}^{\text {inst }}(f)}{\exp \left(-\kappa T_{\text {obs }}^{-1} \frac{d N}{d f}\right)}, S_{h}^{\text {inst }}(f)+S_{h}^{\text {gal }}(f) \mathcal{F}(f)\right]\\
		&+S_{h}^{\text {exgal }}(f) \mathcal{F}(f)+ 0.01 \times S_{h}^{N S}(f),
	\end{aligned}
\end{equation}
where $\kappa \simeq 4.5$, $T_{\text {obs }}$ is the observation time and $dN/ df$ is the number density of white dwarfs in the galactic per unit frequency:
\begin{equation}
	\begin{aligned}
		\frac{d N}{d f}=2 \times 10^{-3}\left(\frac{f}{1 H z}\right)^{-11 / 3} H z^{-1}.
	\end{aligned}
\end{equation}

The factor of 0.01 in $S_{h}^{N S}(f)$ represents the proportion of neutron star noise that cannot be removed. The sensitive frequency range of DECIGO is from $f_{\text{min }}=10^{-3} Hz$ to $f_{\text {max}}=100 Hz$. In Fig.~1, we show the noise spectral density of different GW detectors, from which one could see that gravitational waves generated by binary neutron stars falls within the observation range of DECIGO instead of LISA. 

Based on the one-sided noise power spectral density (PSD), we could obtain the time-domain noise signal from the one-sided PSD; in this paper, we utilized the Python function pycbc.noise.gaussian.noise\_from\_psd \cite{74} ,which takes a PSD as input and returns colored Gaussian noise, to simulate the time-domain noise received by the DECIGO detector.

\begin{figure}[ht]
	\centering
	\includegraphics[width=0.5\textwidth]{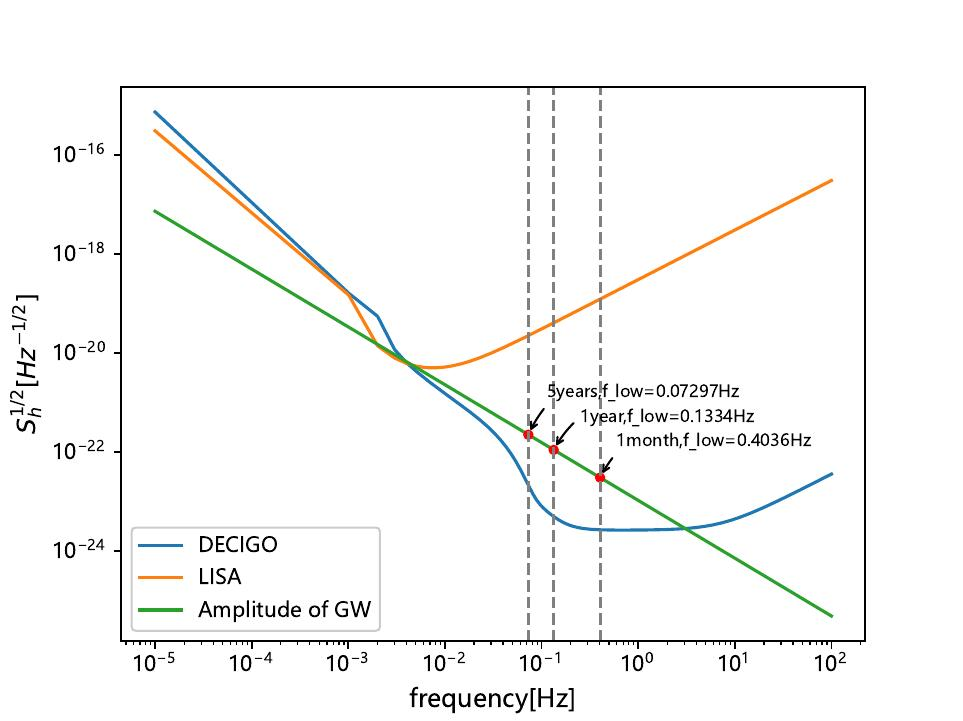}    
	\caption{Sensitivity curves for DECIGO (blue line) and LISA (orange line), as well as the gravitational wave amplitude of a binary neutron star (1.4+1.4) $M_{\odot}$ at a luminosity distance of  $D_{L}=3 G p c$  and a frequency range of  $10^{-5} Hz \sim 100 Hz$  (green line). The arrows on the graph indicate that when observing at the same high frequency of $1 Hz$, the lowest observational frequency differs depending on the observation time. Specifically, for an observation time of $T_{obs}=5 years$, the lowest observational frequency is $f_{\text{min}}=0.073 Hz$. For an observation time of $T_{\text {obs }}=1 year$, the lowest observational frequency is $f_{\text{min}}=0.133 Hz$, and for an observation time of $T_{\text {obs}}=1 month$, the lowest observational frequency is $f_{\text{min}}=0.4036 Hz$.}
	\label{1}
\end{figure}

\subsection{Distribution of BNS and numerical settings}

For the probability density function of the distribution of neutron stars, we adopted the following form \cite{18}:
\begin{equation}
	\begin{aligned}
		\rho(z) \sim \frac{4 \pi d_{C}^{2}(z) R(z)}{H(z)(1+z)},
	\end{aligned}
\end{equation}
where the co-moving distance is $d_{C}(z) = \int_{0}^{z} 1 / H\left(z^{\prime}\right) d z^{\prime}$ and the evolution of the inflation rate with time is quantified as \cite{19,20,21}:
\begin{equation}
	\begin{aligned}
		R(z)=\left\{\begin{array}{cc}
			1+2 z, & z \leq 1 \\
			\frac{3}{4}(5-z), & 1<z<5, \\
			0, & z \geq 5
		\end{array}\right.
	\end{aligned}
\end{equation}
the above probability density function is normalized as: 
\begin{equation}
	\begin{aligned}
		\rho(z)=\frac{4 \pi a d_{C}^{2}(z) R(z)}{H(z)(1+z)},
	\end{aligned}
\end{equation}
with a normalization factor of $a=\frac{1}{\int_{0}^{2} \rho(z)dz}=37502.53$. The distribution function of redshift is given by:
\begin{equation}
	\begin{aligned}
		P(z)=\int_{0}^{z^{\prime}} \rho(z) d z^{\prime},
	\end{aligned}
\end{equation}	
with the probability density function of redshift $\rho(z)$ and the distribution function  $P(z)$ are shown in Fig.~2 (a).  We divide the redshift range into 20 intervals according to the distribution function $P(z)$ and randomly select 1000 redshift values that fall into these intervals. The resulting distribution of 1000 GW sources, based on the distribution function $P(z)$ are presented in Fig.~2 (b).

In our simulation, we set the masses of neutron stars to  $m_{1}=m_{2}=1.4 M_{\odot}$, the merger time to $t_{c}=0$, the coalescence phase $\phi_{c}=0$, and the angles to  $\beta=\sigma=0$. Moreover, we do not include the
effects of the spins of the merging bodies ($S=0$). The high and low frequencies of the gravitational waves and noise are determined as $f_{\text {fin }}=\left(f_{\text {ISCO }}, f_{\text {end }}\right)$ and $f_{\text {in }}=\max \left(f_{\text {obs }}, f_{\text {min }}\right)$. Here $f_{\text {ISCO }}=\left(6^{3/2} \pi M\right)^{-1}$, and $f_{\text{end }}$=1 Hz, within the sensitive range of the DECIGO detector. $f_{\text{obs }}=(256 / 5)^{-3/8} \pi^{-1} \mathcal{M}_{z}^{-5/8} \Delta t_{0}^{-3/8}$ is the lowest frequency that can be observed during the corresponding observation time $\Delta t_{0}=1 month$ \cite{28}. At the luminosity distance of $D_{L}=3 Gpc$, the lowest observational frequency is 0.4036 $Hz$. Therefore, the frequency range of signal and noise is set as 0.4036 $\sim$ 1 Hz.

\begin{figure}[h]
    \centering
    \begin{subfigure}{.5\textwidth}
        \centering
        \includegraphics[width=.9\linewidth]{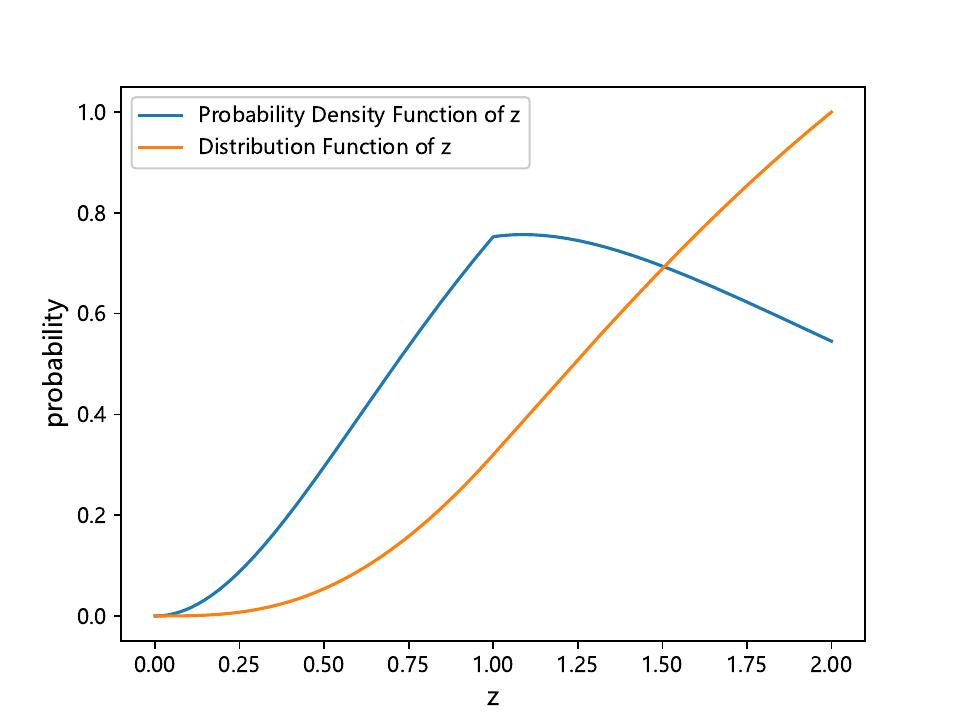}
        \caption{Probability density function and distribution function of redshift.}
    \end{subfigure}
    \begin{subfigure}{.5\textwidth}
        \centering
        \includegraphics[width=.9\linewidth]{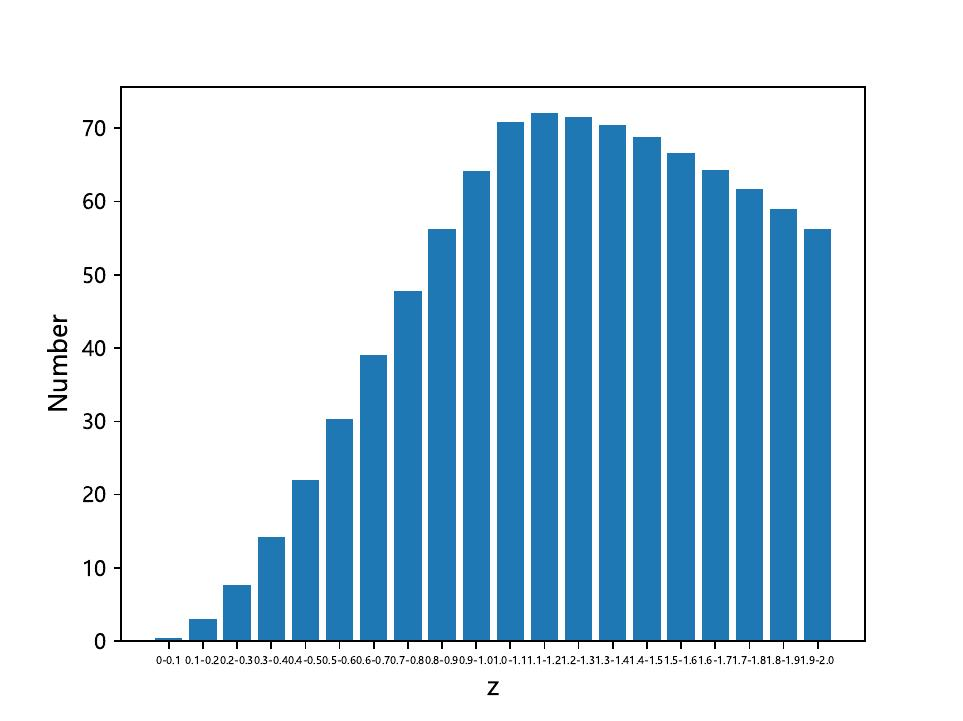}
        \caption{Redshift distribution of 1000 sources. The x-axis and y-axis denote the range of each interval and the number of GW sources in each interval.}
    \end{subfigure}
    \caption{Redshift distribution function and the samples generated according to this distribution.}
\end{figure}

\begin{figure}[ht]
	\centering
	\includegraphics[width=0.5\textwidth]{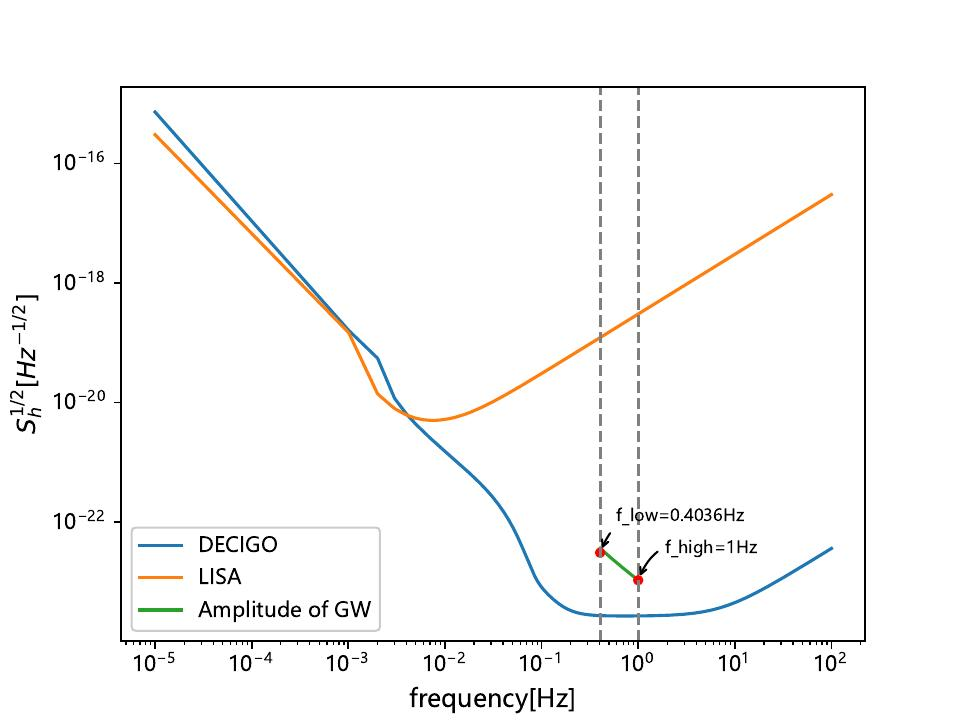}    
	\caption{Frequency-domain amplitude of gravitational wave signal observed for 1 month (yellow line) and one-sided noise power spectral density of DECIGO (blue line), with a frequency range of 0.4036$\sim$1 $Hz$. The binary neutron star with the mass of (1.4+1.4) $M_{\odot}$  is located at a luminosity distance of $D_{L}=3 Gpc$.}
	\label{4}
\end{figure}

\begin{figure}[h]
\centering 
\begin{minipage}{.5\textwidth} 
\centering 
\includegraphics[width=.9\linewidth]{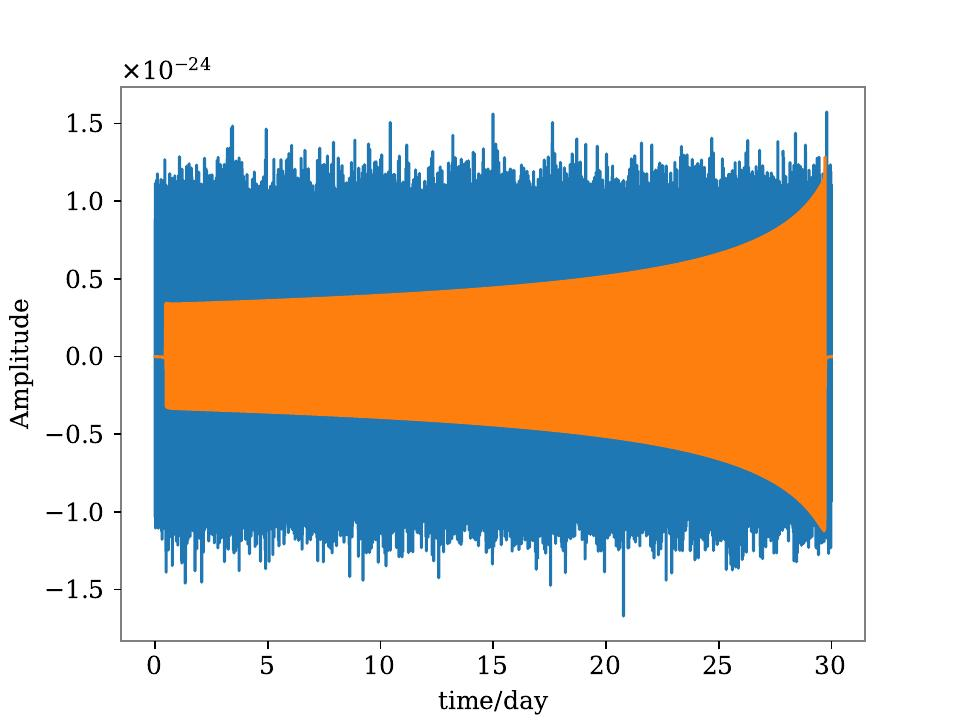} 
\subcaption{ A time-domain data sample.}  \includegraphics[width=.9\linewidth]{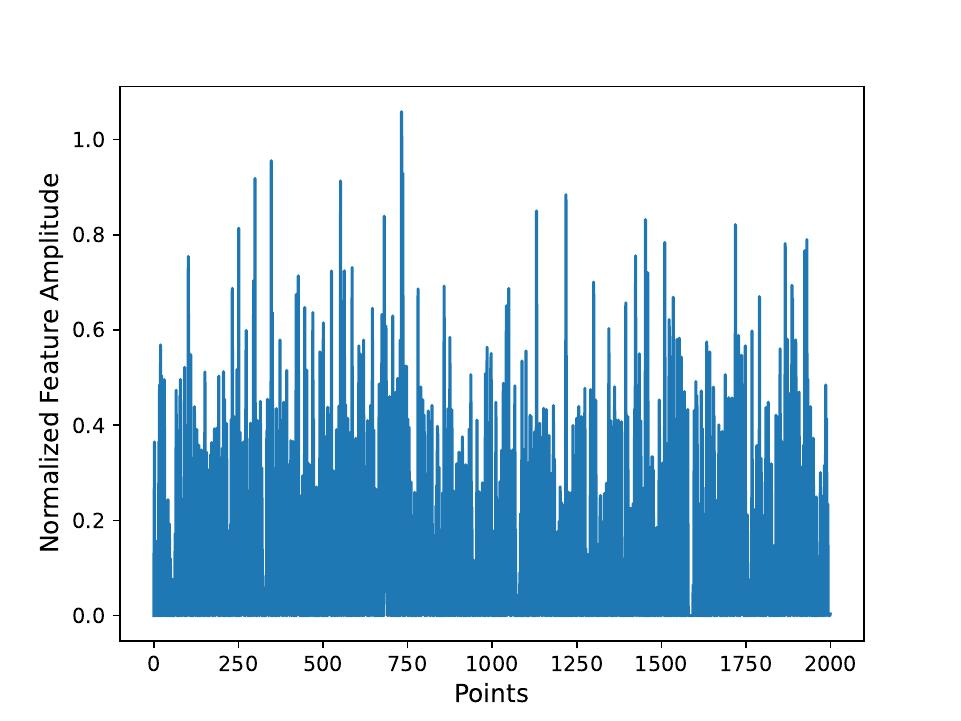} 
\subcaption{The corresponding feature.} 
\end{minipage}% 
% \captionsetup{labelformat=empty,labelsep=none} 
\caption{Noise represented as the blue curve represents, while the yellow curve shows the gravitational wave (a). The sample's S/N is 15.739. The observation time for the signal is 1 month, with a frequency range of  0.4036 $Hz-1 Hz$. The gravitational wave source is a binary neutron star system with a combined mass of  (1.4+1.4) $M_{\odot}$  and a luminosity distance of  $D_{L}=3 Gpc$. Other parameters are set as described in Section 2.4. Here (b) is the corresponding feature of (a), obtained through a single-layer convolutional neural network convoluting with raw data. The horizontal axis of the (b) represents the data length, while the vertical axis represents the normalized feature amplitude.} 
\end{figure}

\section{Estimating the acceleration parameters with deep learning}

\subsection{Construction of the dataset}

We generated 1000 samples of gravitational wave data covering the redshift range of $0<z<2$. The actual input to the network is the feature amplitude extracted from the time-domain data $s(t)=h(t)+n(t)$, using a single-layer convolutional neural network, where $h(t)$ is the inverse Fourier transform of $\tilde{h}(f)$ and $n(t)$ is the time-domain noise. The sampling frequency of the data is taken as 2 Hz, the sampling time is one month and the data length is 1$\times 5184000$. 
However, due to the observation time, resulting in an excessive data length, the total size of our dataset comprising 1000 samples amounts to 38.624 gigabytes(GBs). That leads to significant challenges for training neural networks. Therefore, we employed a 1D convolutional neural network with a single layer for feature extraction from our raw data. Table 1 represents the feature extraction network architecture. Following feature extraction, each sample has a length of 1$\times 2000$. In Fig.~4 (b), we illustrate one sample after feature extraction. We divide the 1000 samples into training and testing dataset, and our testing sub-set contains 30\% of the original sample size. Then we deploy different machine learning algorithms on the training test. The input data is (x\_train, y\_train), (x\_test, y\_test), where x\_train and x\_test are time-domain data in the training and testing datasets. y\_train and y\_test are the  parameters space here is ({$\ln \mathcal{A}, t_c, \phi_c,  \ln \mathcal{M}, \ln \eta, \ln \mathcal{M}_z, X_H$}) that the neural network needs as labels. $X_H = \frac{X(z)}{H_0}$. The S/N values of these 1000 data is shown in Fig.~6.

\begin{table*}[htbp]
	\centering
	\caption{Feature extraction network architecture.}
\begin{tabular}{llll}
\hline
Layer type & Hyper Parameter                                                                                                                    & Input       & Output   \\ \hline
Conv1D     & \begin{tabular}[c]{@{}l@{}}filters=1, kernel\_size=2000, strides=2592,\\ activation='relu',kernel\_initializer='ones'\end{tabular} & (5184000,1) & (2000,1) \\ 
Flatten    &                                                                                                                                    & (2000,1)    & (2000)   \\ \hline
\end{tabular}
\end{table*}

Now our analysis will be carried out on all simulated and data-sets with two deep learning algorithms (CNN and LSTM-GRU), as well as the Fisher information matrix estimation.

\subsection{Convolutional neural networks}

Convolutional neural networks (CNN) offer several advantages for parameter estimation in the analysis of time-domain gravitational wave data. Firstly, a CNN enables automatic feature learning, alleviating the need for manual design of feature extractors \cite{50}. Secondly, it possesses local perception capabilities, allowing it to extract features at different positions within the input data through convolutional filters \cite{52}. This is crucial for capturing local structures and temporal features in time-domain gravitational wave data, facilitating effective discrimination of different gravitational wave signals. Third, CNN employs multiple layers for hierarchical feature representation by stacking convolutional and pooling layers. This progressive learning enables the network to capture abstract features in time-domain gravitational wave data, enhancing the accuracy of parameter estimation. Fourth, CNN exhibits robustness and generalization abilities, acquired through training on large-scale datasets. It demonstrates resilience against noise and non-ideal conditions, which is vital for accurate parameter estimation in the presence of noise and interference in real detectors \cite{51}. Lastly, CNN excels in handling large-scale datasets, a critical advantage for processing the extensive gravitational wave data \cite{53} expected from the next-generation DECIGO detector. As DECIGO is planned to produce high-temporal-resolution data, efficient data processing methods are essential and CNN is capable of effectively managing large-scale data. Collectively, these advantages establish CNN as a powerful tool for parameter estimation in analysis of DECIGO's time-domain gravitational wave data. Based on these characteristics, we chose a 1D CNN model for parameter estimation of gravitational wave signals, with our network structure shown in Table 2.

\begin{table*}[htbp]
	\centering
	\caption{CNN architecture and hyperparameter settings}
	\label{tab:hyperparameters}
	\begin{tabular}{lllllll}
		\hline
		input    & Layer        & size         & Kernel\_number        & Stride & Output    & Parameters \\
		2000*1   & Dense(128)   &              &                       &        & 2000*128  & 256        \\
		2000*50  & Conv1D       & 1*1          & 128                   & 1      & 2000*128  & 6528       \\
		2000*128 & MaxPooling1D & 2*1          &                       & 1      & 1000*128  & 0          \\
		1000*128 & Droupt       & 20\%         &                       &        & 1000*128  & 0          \\
		1000*128 & Conv1D       & 8*1          & 64                    & 1      & 993*64    & 65600      \\
		993*64   & MaxPooling1D & 2*1          &                       & 1      & 496*64    & 0          \\
		496*64   & Droupt       & 20\%         &                       &        & 496*64    & 0          \\
		496*64   & Conv1D       & 8*1          & 32                    & 1      & 489*32    & 16416      \\
		489*32   & Droupt       & 20\%         &                       &        & 489*32    & 0          \\
		489*32   & Conv1D       & 8*1          & 16                    & 1      & 482*16    & 4112       \\
		482*16   & Flatten      &              &                       &        & 7712      & 0          \\
		7712   & Dense(100)   &              &                       &        & 100       & 771300     \\
		100   & Dense(7)     &              &                       &        & 7         & 707        \\
		\hline
		&              & \multicolumn{2}{l}{Loss}             &        & Mean squared error       &            \\
		&              & \multicolumn{2}{l}{Optimizer (Adam)} &        & Lr=0.0003 &            \\
		&              & \multicolumn{2}{l}{Epoch}            &        & 100       &            \\
		&              & \multicolumn{2}{l}{Batch   size}     &        & 16         & \\  
		\hline        
	\end{tabular}
\end{table*}

\subsection{LSTM-GRU Hybrid Network}

The LSTM-GRU hybrid network provides numerous advantages for parameter estimation of time-domain gravitational wave data obtained from the DECIGO detector. Specifically designed to handle sequential data, Long Short-Term Memory (LSTM) and Gated Recurrent Unit (GRU) are variants of recurrent neural networks (RNN) that effectively capture the long-term dependencies often observed in DECIGO gravitational wave data. The LSTM-GRU hybrid network demonstrates robustness and generalization capabilities, making it resilient against noise and incomplete training data \cite{54,55}. Given that DECIGO gravitational wave data may contain measurement errors or missing information, the hybrid network leverages its gating mechanisms and memory units to capture crucial features, enabling reliable parameter estimations even in the presence of such challenges. Additionally, the network exhibits strong generalization ability, allowing accurate estimation of parameters for unseen data samples. Moreover, the LSTM-GRU hybrid network excels in capturing both local and global information. While LSTM captures long-term dependencies by utilizing gating mechanisms and memory cells, GRU can swiftly capture short-term local patterns through its update and reset gates. This combination facilitates comprehensive understanding and modeling of DECIGO time-domain gravitational wave data by enabling the network to focus on both local and global features. Furthermore, The LSTM-GRU hybrid network's nonlinear activation functions and gating mechanisms contribute to its powerful nonlinear modeling capabilities \cite{56}. Traditional linear models often fail to accurately capture the complex nonlinear relationships present in time-domain gravitational wave data. In contrast, the LSTM-GRU hybrid network, with its flexibility in capturing nonlinear patterns and complex data features, enhances the accuracy of parameter estimation. In conclusion, the LSTM-GRU hybrid network offers a range of advantages, including long-term dependency modeling, long-term memory and forgetting capabilities, robustness and generalization capabilities, the ability to capture both local and global information, and powerful nonlinear modeling capabilities. These attributes make the LSTM-GRU hybrid network a valuable tool for parameter estimation of time-domain gravitational wave data acquired from the DECIGO detector. In this paper, we use LSTM and GRU to construct a hybrid model for parameter estimation and prediction. Our LSTM-GRU hybrid network structure is shown in Table 3.

We present the results (under the assumption of unbiased estimation) obtained from the two neural networks in Fig.~5. By comparing the two deep learning methods, we find that the two neural networks perform similarly in terms of parameter estimation in time domain data. However, due to the fewer trainable parameters in the CNN, the training speed of CNNs is faster. For the task of accelerating parameter estimation with the same dataset, the CNN is more efficient than LSTM-GRU.

\begin{table}[htbp]
	\centering
	\caption{Hyperparameters of LSTM-GRU hybrid network}
	\begin{tabular}{lll}
		\hline  
		Layer   type & Hyper   Parameter   & Value     \\
		& Units               & 2000      \\
		LSTM         & Activation   (ReLu) &           \\
		& Dropout             & 0.1       \\
		& Units               & 1000      \\
		GRU          & Activation   (ReLu) &           \\
		& Dropout             & 0.2       \\
		& Units               & 1000      \\
		LSTM         & Activation   (ReLu) &           \\
		& Dropout             & 0.2       \\
		Dense        &                     & 7         \\
		\hline  
		& Loss                & Mean squared error       \\
		& Optimizer (Adam)    & Lr=0.0001 \\
		& Epoch               & 40        \\
		& Batch size          & 16       \\
		\hline  
	\end{tabular}
\end{table}

\begin{figure}[ht]
	\centering
	\includegraphics[width=0.5\textwidth]{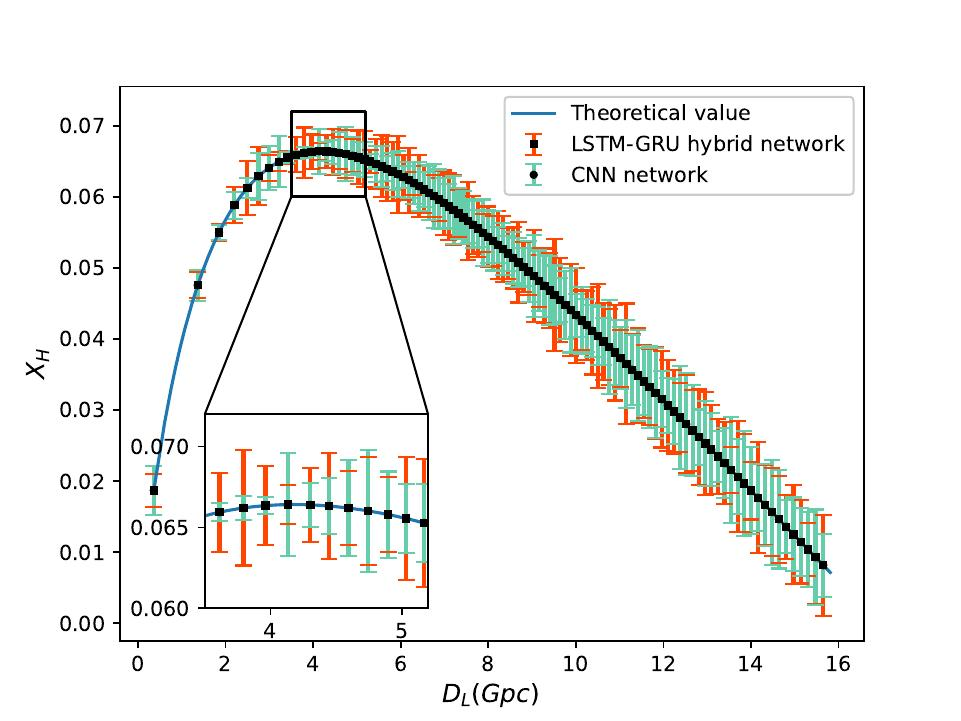}    
	\caption{Comparison of $X_H$ error estimation using CNNs and LSTM-GRU hybrid networks, for binary neutron stars with one month of observations.}
	\label{13}
\end{figure}

\begin{figure}[ht]
	\centering
	\includegraphics[width=0.5\textwidth]{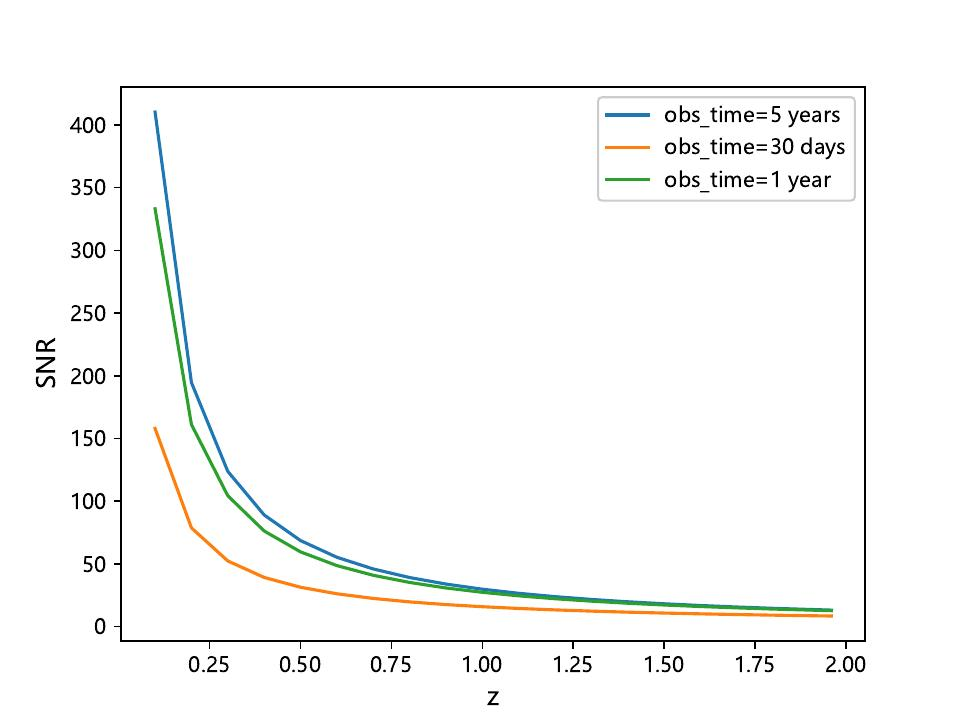}    
	\caption{Variation in the S/N with different observation time. The blue line represents a 5-year observation, the green line represents a 1-year observation, and the yellow line represents a 1-month observation. }
	\label{7}
\end{figure}

\section{Estimating the acceleration parameter with a Fisher information matrix}

In this section, we describe our use of the Fisher information matrix(FIM) for cosmic acceleration parameter estimation. It is based on the methodology of matched filtering to detect gravitational wave signals using fixed templates. The methodology was firstly proposed in \cite{22,23}. Matched filtering convolves the detector output signal with the gravitational wave template in the library to find the template with the maximum correlation. The parameter information of the detection signal is given by the template with the maximum correlation. However, typically the parameter information obtained for the detection signal is not accurate. If the background noise of the signal is Gaussian, the estimated parameters of the signal will have a Gaussian distribution around the actual parameter values.

We assume that $\lambda^{i}$ denotes the actual value of the parameter, $\lambda^{i}+\Delta \lambda^{i}$ denotes the measured value, and the root-mean-square of $\Delta \lambda^{i}$ follows a Gaussian distribution: $p\left(\Delta \lambda^{i}\right) \propto \exp \left(-\Gamma_{i j} \Delta \lambda^{i} \Delta \lambda^{j} / 2\right)$. Here $\Gamma_{i j}$ is the matrix element of the Fisher information matrix:
\begin{equation}
	\begin{aligned}
		\Gamma_{i j}=2 \int_{f_{i n}}^{f_{f i n}} \frac{\tilde{h}_{i}^{*}(f) \tilde{h}_{j}(f)+\tilde{h}_{i}(f) \tilde{h}_{j}^{*}(f)}{S_{h}(f)} d f,
	\end{aligned}
\end{equation}
where $\tilde{h}_{i}(f) = \partial \tilde{h}(f) / \partial \lambda^i$, $\tilde{h}(f)$ is the frequency-domain form of the gravitational wave signal, and $\tilde{h}^*(f)$ is the complex conjugate of $\tilde{h}(f)$. 
The root-mean-square error of the $i$-th parameter is $\sigma^i = \sqrt{\Sigma^{ii}}$, where $\Sigma^{ii} = [\Gamma^{-1}]^{ii}$ is the covariance matrix element for parameter $i$. The root-mean-square error of the $i$-th parameter is $\sigma^i = \sqrt{\Sigma^{ii}}$. The off-diagonal elements in the $\Sigma$ matrix represent the size of the correlation between the $i$-th and $j$-th parameters, denoted as $c^{ij}$ and calculated as $c^{ij} = \frac{\Sigma^{ij}}{\sqrt{\Sigma^{ii}\Sigma^{jj}}}$. The parameter space here is ({$\ln \mathcal{A}, t_c, \phi_c,  \ln \mathcal{M}, \ln \eta, \ln \mathcal{M}_z, X_H$}), where $X_H = \frac{X(z)}{H_0}$, and all parameter settings are given in Section 2. The S/N is calculated as:
\begin{equation}
	\begin{aligned}
		SNR=4 \times N_{\text {int }} \int_{f_{\text {in }}}^{f_{f i n}} \frac{|\tilde{h}(f)|^{2}}{S_{n}(f)} d f,
	\end{aligned}
\end{equation}
where $N_{\text {int}}=8$ represents the number of effective interferometer arms of DECIGO. The S/N values for observation times of 1 month and 5 years are shown in Fig.~6.

It should be noted that the results obtained by our neural networks are based on the assumption of unbiased estimation. Here, $\Delta X_H$ represents the standard deviation of $X_H$ estimated by our model. For our neural networks model, we performed 1000 predictions, resulting in 1000 samples of $X_H$ at each redshift. Subsequently, the standard deviation $\Delta X_H$ was computed using Eq.~21. In contrast, for our Fisher method, we directly derived the standard deviation by calculating the covariance matrix and taking the square root of its diagonal elements.

\begin{equation}
	\begin{aligned}
        \Delta X_{H}=\sqrt{\frac{\sum_{i=1}^{n}\left(x_{i}-X_{H}\right)^{2}}{n}},
	\end{aligned}
\end{equation}
where $x_{i}$ represents our samples,n=1000, which is the size of samples.

Given a one month of observations with DECIGO, the estimated results of $\Delta X_H$ are shown in Table 4. We further studied the estimated uncertainty $\Delta X_{H}$ with 5-year observation of DECIGO. The results in Table 5 are well consistent with those obtained in the previous works \cite{10}. Based on five years of observation, we can essentially estimate the acceleration parameter with reasonable uncertainties, allowing for a rough assessment of the cosmic accelerated expansion. Our finding indicate that increasing the observation time can improve the S/N and thus reduce the uncertainties of $X_{H}$. Such a conclusion is strongly supported by the results shown in Fig.~7, from which we may also find that the estimated error gradually increases with the luminosity distance.

Finally, Fig.~8 shows the results of both methods. We can also calculate the relative error $\frac{\Delta X_{H}}{X_{H}} \times 100\%$ and visualize the results in Fig.~9. Based on the simulated gravitational wave data with a time duration of 1 month, the CNN can limit the relative error to 15.71\%, while the LSTM network combined with GRU can limit the relative error to 14.14\%. Additionally, using Fisher information matrix for gravitational wave data with a 5-year observation can limit the relative error to 32.94\%.
Therefore,the neural networks can give a high-precision estimation of the acceleration parameter at different redshifts. In this case, DECIGO is expected to provide direct measurements of the acceleration of the universe, by observing the chirp signals of coalescing binary neutron stars.

\begin{table}[htbp]
	\centering
	\caption{Estimation of $\Delta X_{H}$ with 1-month observation of binary neutron stars at different redshifts, based on Fisher information matrix. }
	\label{1}
	\begin{tabular}{c c c c c }
		\hline
		$z$ & $f_{in} (Hz)$ & SNR & $X_H$ & $\Delta X_H$ \\ \hline
		0.1 & 0.495 & 157.907 & 0.0228 & 0.0283 \\ 
		0.2 & 0.469 & 78.618 & 0.0389 & 0.1058 \\ 
		0.3 & 0.445 & 52.243 & 0.0500 & 0.1802 \\ 
		0.4 & 0.425 & 39.101 & 0.0572 & 0.3747 \\ 
		0.5 & 0.407 & 31.250 & 0.0613 & 0.2132 \\ 
		0.6 & 0.391 & 26.042 & 0.0632 & 0.4431 \\ 
		0.7 & 0.377 & 22.341 & 0.0632 & 0.3530 \\ 
		0.8 & 0.363 & 19.579 & 0.0618 & 0.5228 \\ 
		0.9 & 0.351 & 17.441 & 0.0592 & 0.0660 \\ 
		1.0 & 0.340 & 15.739 & 0.0558 & 0.2683 \\ 
		1.1 & 0.330 & 14.351 & 0.0517 & 0.7306 \\ 
		1.2 & 0.320 & 13.199 & 0.0470 & 0.8172 \\ 
		1.3 & 0.312 & 12.227 & 0.0419 & 0.8096 \\ 
		1.4 & 0.303 & 11.395 & 0.0364 & 1.4532 \\ 
		1.5 & 0.296 & 10.676 & 0.0307 & 0.7816 \\ 
		1.6 & 0.289 & 10.047 & 0.0248 & 1.9325\\ 
		1.7 & 0.282 & 9.493 & 0.0187 & 1.4471 \\ 
		1.8 & 0.275 & 9.001 & 0.0124 & 1.0913 \\ 
		1.96 & 0.266 & 8.318 & 0.0080 & 2.0143 \\ 
		\hline
	\end{tabular}
\end{table}

\begin{table}[htbp]
	\centering
	\caption{Estimation of $\Delta X_{H}$ with 5-year observation of  binary neutron stars at different redshifts, based on Fisher information matrix.}
	\label{2}
	\begin{tabular}{c c c c c}
		\hline
		$z$ & $f_{in} (Hz)$ & SNR & $X_H$ & $\Delta X_H$ \\ \hline
		0.1 & 0.106 & 410.330 & 0.0228 & 0.0002 \\ 
		0.2 & 0.100 & 194.515 & 0.0389 & 0.0007 \\ 
		0.3 & 0.095 & 123.686 & 0.0500 & 0.0013 \\ 
		0.4 & 0.091 & 88.931 & 0.0572 & 0.0021 \\ 
		0.5 & 0.087 & 68.503 & 0.0613 & 0.0029 \\ 
		0.6 & 0.083 & 55.171 & 0.0632 & 0.0038 \\ 
		0.7 & 0.080 & 45.851 & 0.0632 & 0.0048 \\ 
		0.8 & 0.077 & 39.010 & 0.0618 & 0.0059 \\ 
		0.9 & 0.075 & 33.800 & 0.0592 & 0.0070  \\ 
		1.0 & 0.072 & 29.717 & 0.0558 & 0.0082 \\ 
		1.1 & 0.070 & 26.442 & 0.0517 & 0.0094 \\ 
		1.2 & 0.068 & 23.764 & 0.0470 & 0.0105 \\ 
		1.3 & 0.066 & 21.540 & 0.0419 & 0.0116  \\ 
		1.4 & 0.065 & 19.667 & 0.0364 & 0.0125  \\ 
		1.5 & 0.063 & 18.070 & 0.0307 & 0.01338  \\ 
		1.6 & 0.061 & 16.695 & 0.0248 & 0.0138\\ 
		1.7 & 0.060 & 15.501 & 0.0187 & 0.0139 \\ 
		1.8 & 0.059 & 14.454 & 0.0124 & 0.0138 \\ 
		1.96 & 0.057 & 13.026 & 0.0080 & 0.0140 \\ \hline
		
	\end{tabular}
\end{table}

\begin{figure}[htbp]
	\centering
	\includegraphics[width=0.5\textwidth]{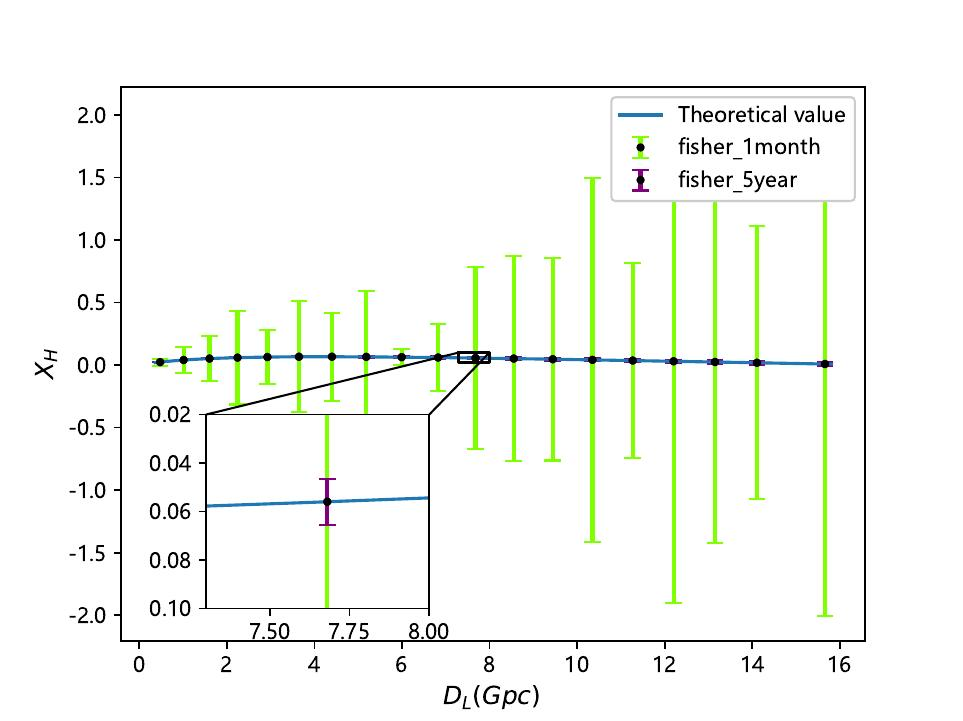}    
	\caption{Comparison of $X_H$ estimation from 1-month and 5-year observations of binary neutron stars, based on Fisher information matrix.}
	\label{10}
\end{figure}

\begin{figure}[h]
\centering 
\begin{minipage}{.5\textwidth} 
\centering 
\includegraphics[width=.9\linewidth]{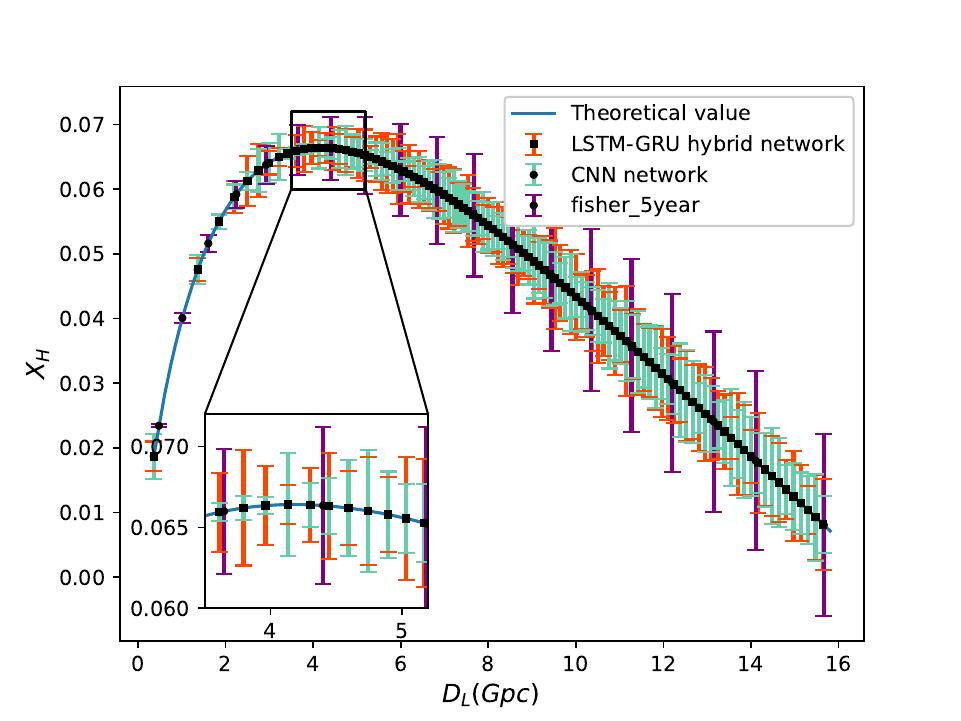} 
\subcaption{$X_H$ estimation using deep learning with 1-month observation and Fisher information matrix with 5-year observation.}  \includegraphics[width=.9\linewidth]{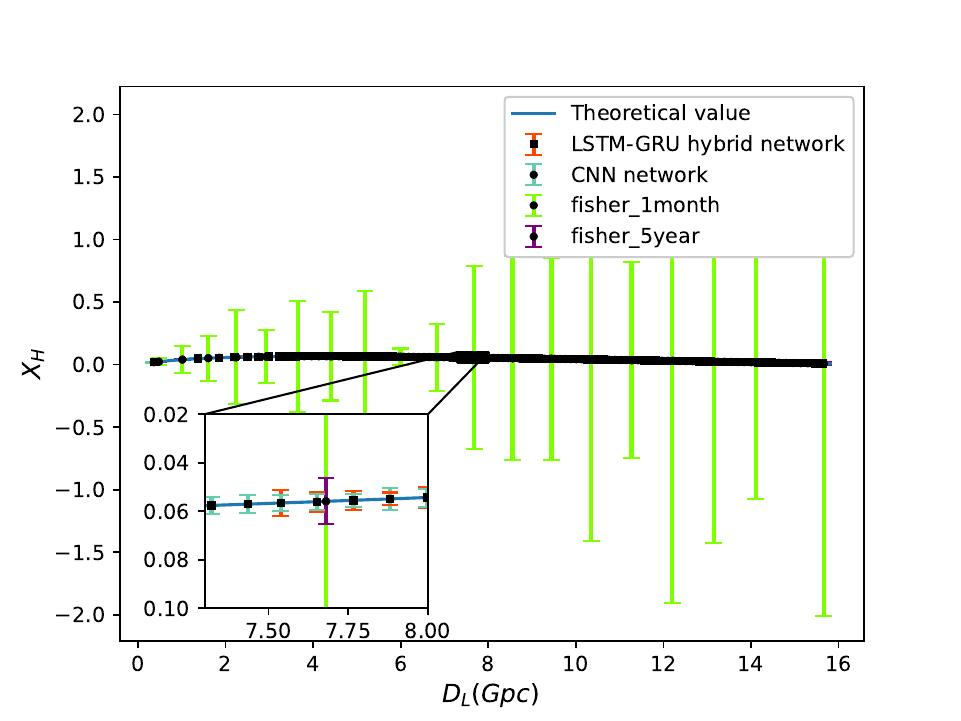} 
\subcaption{$X_H$ estimation using deep learning with 1-month observation and Fisher information matrix with 1-month and 5-year observation.} 
\end{minipage} 
\caption{$X_H$ estimation using deep learning and Fisher information matrix} 
\end{figure}

\begin{figure}[ht]
	\centering
	\includegraphics[width=0.5\textwidth]{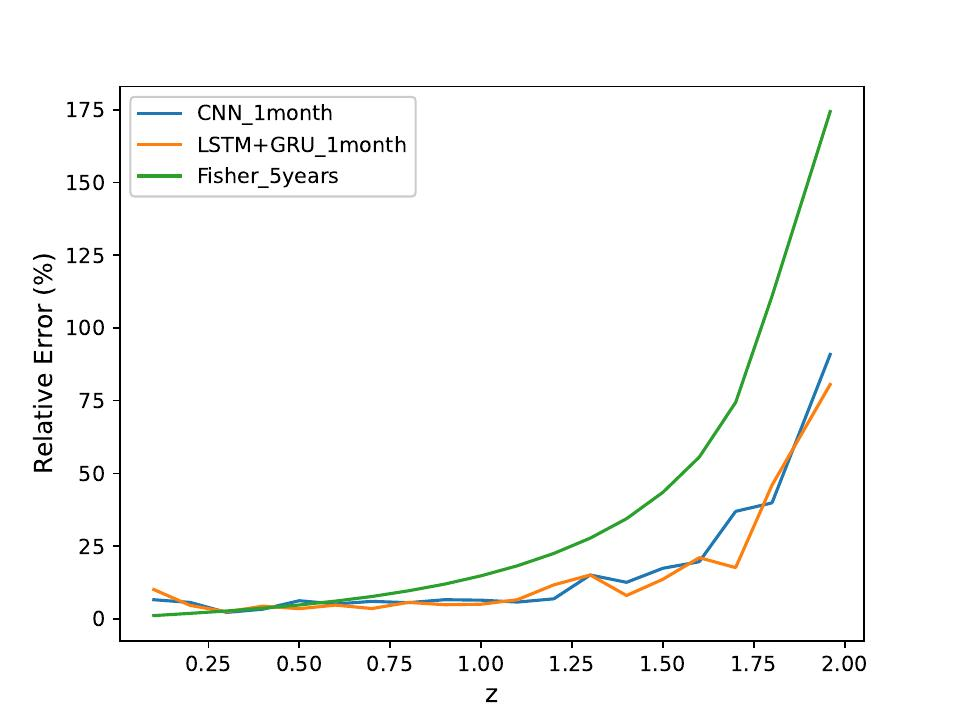}    
	\caption{Error estimated by the Fisher matrix for 5-year observation data, shown as a green solid line. The orange solid line represents the error value estimated by the LSTM-GRU networks for one month of observation data, while the blue solid line represents the error value estimated by the CNN networks for one month of observation data.}
	\label{16}
\end{figure}

\begin{table}[htbp]
	\centering
	\caption{Estimation of $\Delta X_{H}$ using Fisher information matrix with a 5-year observation and deep learning with a 1-month observation.}
	\begin{tabular}{c c c c c}
		\hline
		z & $X_H$ & $\Delta X_H(\text{Fisher})$ & $\Delta X_H(\text{CNN})$ & $\Delta X_H(\text{L-G})$ \\
		\hline
		0.1 & 0.0228 & 0.0002 & 0.0015 & 0.0023 \\
		0.2 & 0.0389 & 0.0007 & 0.0022 & 0.0018 \\
		0.3 & 0.0500 & 0.0013 & 0.0011 & 0.0012 \\
		0.4 & 0.0572 & 0.0020 & 0.0019 & 0.0025 \\
		0.5 & 0.0613 & 0.0029 & 0.0031 & 0.0012 \\
		0.6 & 0.0632 & 0.0038 & 0.0026 & 0.0022 \\
		0.7 & 0.0632 & 0.0048 & 0.0053 & 0.0015 \\
		0.8 & 0.0618 & 0.0059 & 0.0029 & 0.0036 \\
		0.9 & 0.0592 & 0.0070 & 0.0021 & 0.0023 \\
		1.0 & 0.0558 & 0.0082 & 0.0041 & 0.0030 \\
		1.1 & 0.0517 & 0.0094 & 0.0025 & 0.0036 \\
		1.2 & 0.0470 & 0.0105 & 0.0036 & 0.0028 \\
		1.3 & 0.0419 & 0.0116 & 0.0060 & 0.0047 \\
		1.4 & 0.0364 & 0.0125 & 0.0035 & 0.0030 \\
		1.5 & 0.0307 & 0.0133 & 0.0038 & 0.0052 \\
		1.6 & 0.0248 & 0.0138 & 0.0046 & 0.0068 \\
		1.7 & 0.0187 & 0.0139 & 0.0043 & 0.0026 \\
		1.8 & 0.0124 & 0.0138 & 0.0046 & 0.0057 \\
		1.96 & 0.0023 & 0.0140 & 0.0069 & 0.0073 \\
		\hline
	\end{tabular}
\end{table}

\section{Summary and discussion}

In this paper, we explore the possibility of constraining the cosmic acceleration parameters with the inspiral gravitational waveform of neutron star binaries (NSBs) in the frequency range of 0.1Hz-10Hz, which can be detected by the second-generattion space-based gravitational wave detector DECIGO. We use a convolutional neural network (CNN), a long short-term memory (LSTM) network combined with a gated recurrent unit (GRU), and Fisher information matrix to derive constraints on the cosmic acceleration parameter,  $X_H$. Under the assumption of unbiased estimation, based on the simulated gravitational wave data with a time duration of 1 month, we conclude that the CNN can limit the relative error to 15.71\%, while the LSTM network combined with GRU can limit the relative error to 14.14\%. Additionally, using Fisher information matrix for gravitational wave data with a 5-year observation can limit the relative error to 32.94\%. Therefore, DECIGO is expected to provide an unprecedented opportunity for high-precision detection of cosmic acceleration, by observing the chirp signals of coalescing binary neutron stars \cite{26,60}. 

We should stress that the present paper is only an interesting example of extensive applications of deep learning in cosmological studies \cite{57,58,59}. Still, there are several remarks that remain to be clarified as follows. Firstly,
the deep learning models used can be further optimized and enhanced to improve the measurements of cosmic acceleration parameters. We can explore the use of other neural networks or combinations of multiple networks to achieve more stringent cosmological constraints \cite{61,62,63}. Combining deep learning methods with other approaches, such as Markov Chain Monte Carlo (MCMC) method, could contribute to resolving such an important issue. Secondly, the GW observations provide a powerful and novel method to detect the cosmic acceleration in a cosmological-model-independent way. This strengthens the probative power of such method to inspire new observing programs in the framework of DECIGO, focusing on a large number of neutron-star binaries in inspiraling phases. Finally, in future works, we can apply  deep learning methods to constrain the cosmological parameters associated with the selected scientific objectives encompassed by the DECIGO\cite{64,65,66,67,68,69,70,71,72,73}. This will open up a new window for gravitational-wave cosmology.

\section*{Acknowledgements}

This work is supported by the National Key Research and Development Program of  China (Grant No. 2021YFC2203004); the National Natural Science Foundation of China under Grants Nos. 12021003, 12147102, 11690023, and 11920101003;the Natural Science Foundation of Chongqing (Grant No. CSTB2023NSCQ-MSX0103); the Strategic Priority Research Program of the Chinese Academy of Sciences, Grant No. XDB23000000; and the Interdiscipline Research Funds of Beijing Normal University.

\bibliographystyle{IEEEtran}
\bibliography{Constrain_the_Cosmic_Acceleration_Parameter_with_DECIGO}

\end{document}